\documentclass[aps, prl, twocolumn]{revtex4-1}

\usepackage {graphicx}
\usepackage {amsmath}
\usepackage {amsfonts}
\usepackage {amssymb}
\usepackage {mathrsfs}
\usepackage {bm}

\usepackage {epstopdf}
\allowdisplaybreaks[4]

\newcommand {\be}{\begin{eqnarray}}
\newcommand {\ee}{\end{eqnarray}}

\begin{document}

\title{Theory of Charge Order and Heavy-Electron Formation in the Mixed-Valence Compound KNi$_2$Se$_2$}

% repeat the \author .. \affiliation  etc. as needed
% \email, \thanks, \homepage, \altaffiliation all apply to the current
% author. Explanatory text should go in the []'s, actual e-mail
% address or url should go in the {}'s for \email and \homepage.
% Please use the appropriate macro foreach each type of information

% \affiliation command applies to all authors since the lasta
% \affiliation command. The \affiliation command should follow the
% other information
% \affiliation can be followed by \email, \homepage, \thanks as well.

\author{James M. Murray and Zlatko Te\v{s}anovi\'{c}}
\affiliation{Institute for Quantum Matter and Department of Physics and Astronomy, Johns Hopkins University, Baltimore, MD 21218}

%Collaboration name if desired (requires use of superscriptaddress
%option in \documentclass). \noaffiliation is required (may also be
%used with the \author command).
%\collaboration can be followed by \email, \homepage, \thanks as well.
%\collaboration{}
%\noaffiliation

\date{\today}

\begin{abstract}
The material KNi$_2$Se$_2$ has recently been shown to possess a number of striking physical properties, many of which are apparently related to the mixed valency of this system, in which there is on average one quasi-localized electron per every two Ni sites. Remarkably, the material exhibits a charge density wave (CDW) phase that disappears upon cooling, giving way to a low-temperature coherent phase characterized by an enhanced electron mass, reduced resistivity, and an enlarged unit cell free of structural distortion. Starting from an extended periodic Anderson model and using the slave-boson formulation, we develop a model for this system and study its properties within mean-field theory. We find a reentrant first-order transition from a CDW phase, in which the localized moments form singlet dimers, to a heavy Fermi liquid phase as temperature is lowered. The magnetic susceptibility is Pauli-like in both the high- and low-temperature regions, illustrating the lack of a single-ion Kondo regime and free local moment behavior such as that usually found in heavy-fermion materials.
\end{abstract}

% insert suggested PACS numbers in braces on next line
% \pacs{}
% insert suggested keywords - APS authors don't need to do this
%\keywords{}

%\maketitle must follow title, authors, abstract, \pacs, and \keywords
\maketitle

% body of paper here - Use proper section commands
% References should be done using the \cite, \ref, and \label commands
%\section{}
% Put \label in argument of \section for cross-referencing
%\section{\label{}}
%\subsection{}
%\subsubsection{}

%\section{Introduction}
Heavy-fermion materials exhibit a host of fascinating collective quantum behaviors, which have made them a major focus of ongoing research for over three decades \cite{coleman07, hewson93}. The ``standard model'' of heavy-fermion behavior, as depicted in the famed Doniach diagram \cite{doniach77}, features competition between a magnetic phase and the heavy Fermi liquid. 
%In this picture, the local moments, typically at a density of one moment per atomic site, behave at high temperatures as free spins. Upon cooling, these localized moments may either hybridize with the conduction electrons to form heavy quasiparticles with a large effective mass, or else they may remain localized and order magnetically. Which of these options is realized depends on the parameters of the system, which might be tunable experimentally by applied pressure, doping, or other means. In this way one finds quite generally a competition between the heavy-fermion state and magnetic behavior in these materials.
An intriguing possibility that has received comparatively little attention is the existence of charge order, rather than the usual magnetic order, in proximity to the heavy-fermion state. Mixed-valency systems \cite{varma76}, which contain a variable number of localized electrons per atomic site, are a natural place to look for such competing effects. If the fractional filling takes a commensurate value, then Coulomb repulsion between electrons on nearby sites may induce charge density wave (CDW) ordering. Mixed valency has been studied recently in $f$-electron materials exhibiting heavy-fermion behavior \cite{kummer11, fernandez-panella12, watanabe11}, as well as in the context of the related ``charge Kondo effect'' \cite{taraphder91, matsuura12}. However, in both of those cases the emphasis has generally been on the single-site valency as hybridization or interaction between electrons is increased, rather than on the possibility of collective CDW formation and competition of this charge order with the heavy fermion phase.

%reduced count of localized $f$-electrons on each individual site as hybridization with conduction electrons increases and the $f$-electrons delocalize and participate in the heavy Fermi liquid. 

The material KNi$_2$Se$_2$ has recently been shown to exhibit several remarkable physical properties \cite{neilson12}, many of which appear to be related to its mixed-valent nature. (see also Refs.\ \cite{neilson13,lei12} for recent work on related materials.) At high temperatures the material has high resistivity; the magnetic susceptibility is constant, indicating Pauli paramagnetic response; and structural analysis reveals that the material has at least three distinct sub-populations of Ni-Ni bond lengths. Upon cooling below $T_\mathrm{coh} \approx 20 \mathrm{K}$, the resistivity rapidly decreases, the structural distortions disappear, and the material enters a coherent heavy-fermion state with effective electron mass $m^* \sim 10 m_0$, eventually giving way to superconductivity below $T_c \approx 1 \ \mathrm{K}$. This material is also unusual in that an applied magnetic field induces virtually no response in the measured specific heat and resistivity, indicating that the low-temperature coherent phase does not arise from competition with local magnetic order as in typical heavy-fermion materials. Rather, it was proposed that the coherent state competes with a {\em charge}-fluctuating state, facilitated by the mixed valency of the Ni ions in KNi$_2$Se$_2$ \cite{neilson12}. 
%The proximity of this system to charge (rather than magnetic) order, the disappearance of the CDW as temperature is lowered, and the lack of common signatures of Kondo-like behavior in the magnetic response together distinguish this material from conventional heavy-electron systems, making it a new and fascinating example of the diversity of effects that can be observed in these materials.  

In this study, we present a theory that captures the key ingredients that characterize this system. At high temperatures, the quasi-localized electrons in our model form a CDW and pair with one another into singlet dimers, which explains the observed structural distortion and insensitivity to applied magnetic field. As temperature is lowered, the CDW dissolves in a first-order transition directly into a spatially uniform, correlated heavy-fermion state, without the intermediate single-ion Kondo regime that is usually observed in heavy-fermion materials. The details of this model and the main results of the calculations are presented below. 

%\section{Mean-field theory for the CDW transition}
KNi$_2$Se$_2$ has a quasi-two-dimensional, layered structure, 
%similar to that of heavy fermion materials such as URu$_2$Si$_2$ and the ``122'' family of iron-pnictide superconductors, 
with the Ni and Se ions alternating in checkerboard fashion on a square lattice within each layer. Consideration of the stoichiometry reveals that the effective valency of Ni in this compound is ``$1.5+$,'' so that at low energies the effective degree of freedom is one quasi-localized $d$-electron with spin $1/2$ per every two Ni sites, with a small amplitude for these electrons to hop to neighboring Ni sites. Conduction electron bands are formed from the other Ni and Se orbitals and have a significantly greater bandwidth than the quasi-localized $d$-electrons. 

With this picture in mind, the following Hamiltonian describing the ``extended periodic Anderson model'' provides a useful starting point:
\be
\label{hea}
\begin{split}
H_\mathrm{EA} = &- t_c \sum_{\langle i j \rangle , \sigma} ( c^\dagger_{i \sigma} c_{j \sigma} + H.c.) 
	- t_f \sum_{\left< ij \right> , \sigma } ( d^\dagger_{i \sigma} d_{j \sigma} + H.c.)
	\\ & - \varepsilon_f \sum_i n_{di} + V \sum_{i, \sigma} (d^\dagger_{i \sigma} c_{i \sigma} + H.c.) 
	\\ & + U \sum_{i} n_{di \uparrow} n_{di \downarrow} +  \sum_{i \neq j} W_{ij} n_{di} n_{dj} ,
\end{split}
\ee
where $i,j$ denote Ni sites on a two-dimensional square lattice. The first term in this equation describes hopping of the conduction electrons. The second and third terms describe the hopping and on-site energy of quasi-localized electrons on neighboring Ni sites. The fourth term describes hybridization between the two types of electrons. Finally, the last two terms describe Coulomb repulsion of $d$-electrons occupying the same site and nearby sites, where $n_{di\sigma} = d^\dagger_{i\sigma} d_{i\sigma}$. The Hamiltonian is identical to the well-known periodic Anderson model, with the addition of the $W$ term describing intersite Coulomb repulsion. This term is typically neglected in describing heavy fermion materials since such systems usually have exactly one local moment per site, so such a Coulomb term effectively adds an overall constant to the total energy. It is crucial for describing a system near one quarter filling, however, since such systems are susceptible to Coulomb repulsion-driven charge ordering.

In the limit of large on-site repulsion $U$, it is convenient to enforce the constraint of no double occupancy through the introduction of slave boson operators \cite{read83, coleman84, newns87}. In this formulation, we substitute $d_{i \sigma} = b^\dagger_i f_{i \sigma}$, where $f_{i \sigma}$ describes a charge-neutral ``spinon'' that carries the spin of the electron, and the slave boson operator $b_i$ describes a spinless particle with positive charge. 
%Thus the operation of annihilating a $d$-electron in this language corresponds to removing a neutral spin from site $i$ and replacing it with a hole having positive charge. 
%Within this formalism, rather than having to enforce the constraint $n_{di} \leq 1$ at every site, one instead has the constraint $n_{fi} + b^\dagger_i b_i = 1$, where $n_{fi} = f^\dagger_{i \uparrow} f_{i \uparrow} + f^\dagger_{i \downarrow} f_{i \downarrow}$ is the number operator for spinons. 
In terms of these new operators, the Hamiltonian \eqref{hea} becomes
\begin{align*}
\label{h}
%\begin{split}
H &=  H_c + H_{fc} + H_f  + H_W + H_\lambda + H_J \tag{2}
\\ H_c &= -t_c \sum_{\left< ij \right> , \sigma} (c^\dagger_{i \sigma} c_{j \sigma} + H.c.)
\\ H_{fc} &= V \sum_{i, \sigma} (b_i f^\dagger_{i \sigma} c_{i \sigma} + H.c.)
\\ H_f &= -t_f \sum_{\left< ij \right> , \sigma} ( b_i b^\dagger_j f^\dagger_{i \sigma} f_{j \sigma} + H.c.) 
	- \varepsilon_f \sum_i n_{fi}
\\ H_W &= \sum_{i \neq j} W_{ij} n_{fi} n_{fj}
\\ H_\lambda &= i \sum_i \lambda_i (n_{fi} + b^\dagger_i b_i - 1)
\\ H_J &= J \sum_{\left< ij \right>} {\bf S}_i \cdot {\bf S}_j
%\end{split}
\end{align*}
The first four terms in \eqref{h} are analogous to terms appearing in \eqref{hea}, but rewritten in the slave boson description. $H_\lambda$ replaces the on-site repulsion term in \eqref{hea} by enforcing the constraint  $n_{fi} + b^\dagger_i b_i = 1$ via the Lagrange multiplier field $\lambda_i$ (within mean-field theory, this constraint is enforced only on average). The last term in \eqref{h} describes an antiferromagnetic Heisenberg interaction driven by superexchange between spins on neighboring sites, which is present in the limit where $U$ is large but not infinite. The Hamiltonian \eqref{h} is identical to the ``Anderson--Heisenberg'' model that has been studied recently \cite{pepin07, zhu08, tran12}, with the addition of the Coulomb term $H_W$.
%Because the value of $J$ will be sensitive to details such as the conduction band filling, lattice spacing and other details that we are not particularly concerned with, we shall here include the Heisenberg term as an effective interaction, with the parameter $J$ treated as independent of the other parameters in the model. 
%In principle the RKKY contribution to this term is dynamically generated from the hybridization term in \eqref{hea} upon integrating out the $c$-electrons. In order to study the model within mean-field theory, however, the RKKY contribution is taken to be present in the model as an effective interaction \cite{tran12}.

%$H_c$ describes the hopping of electrons in the conduction band; $H_{fc}$ describes the hybridization between conduction electrons $c_{i \sigma}$ and spinons $f_{i \sigma}$; $H_{f}$ describes the spinon hopping and on-site energy;  $H_J$ describes the antiferromagnetic interaction between spinons; $H_W$ describes Coulomb repulsion between nearest-, second- and third-neighbor spinons; and .

\begin{figure}
\includegraphics[width=0.45\textwidth]{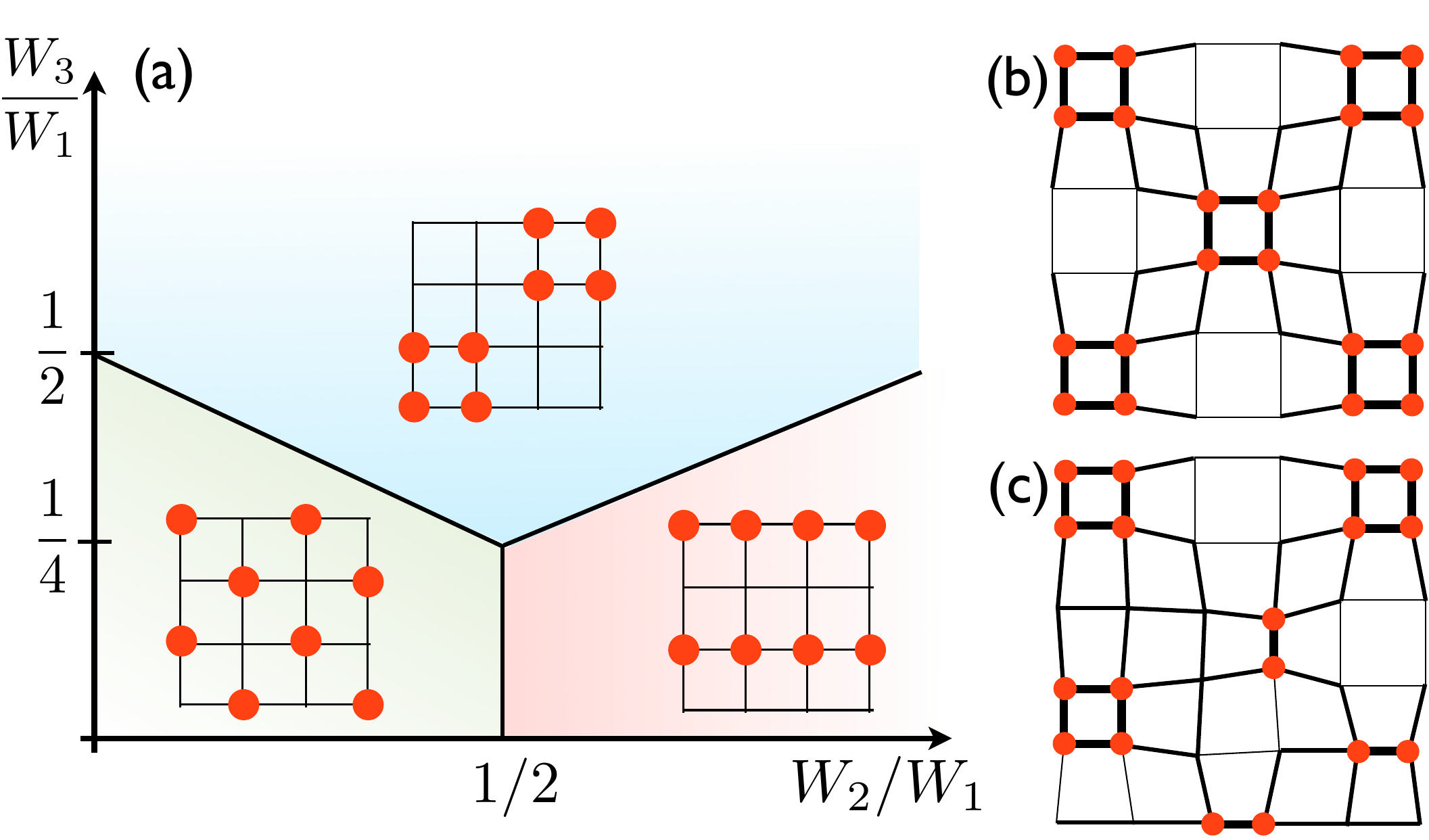}
\caption{(a) $T=0$ diagram of CDW phases minimizing the term $H_W$ for various values of second- and third-neighbor Coulomb repulsion. (b) Schematic illustration of the ``plaquette'' phase showing expected lattice distortion once the coupling between electron and lattice degrees of freedom is taken into account. Links connecting sites containing localized electrons are shorter and are represented by thicker lines. (c) Schematic illustration of a possible configuration in the case where long-range CDW order is absent, but the singlet nature of the state and distribution of bond lengths remain similar to those shown in (b).
\label{cdw_diagram}}
\end{figure}

%Due to the fractional filling of spinons, there is a tendency to develop CDW order when the Coulomb repulsion terms $W_{ij}$ in \eqref{h} are sufficiently large. 
Figure \ref{cdw_diagram}(a) shows a phase diagram with possible CDW phases for various values of first-, second-, and third-nearest neighbor repulsion.
We choose parameters such that the ``plaquette'' CDW phase in the upper part of Figure \ref{cdw_diagram}(a) is realized, since this phase naturally allows for dimer formation between neighboring spinons.
%In order to describe the observed absence of high-termperature Curie magnetic response in KNi$_2$Se$_2$, the model should allow for spinons to form singlet pairs with some of their nearest neighbors. Because the spinons in the phase shown in the lower-left part of Figure \ref{cdw_diagram} have no nearest neighbors, while the phase in the lower-right would lead to quasi-one-dimensional behavior that has not been observed experimentally, the ``plaquette'' charge order at large $W_3$ is chosen as the best candidate. In particular, it is possible for the plaquette phase to be realized for natural choices of couplings where $W_1 > W_2 > W_3$. 
%The realization of this phase does not require especially large values of third-neighbor repulsion $W_3$. 
%For example, the plaquette CDW phase is realized in the idealized case of $\sim 1/r$ Coulomb repulsion, for which the ratios of the repulsion terms are $W_2/W_1 = 1/\sqrt{2}$ and $W_3/W_1 = 1/2$. 
%This plaquette order is also degenerate with the ``double stripe'' order that is realized in the spin-density wave phase of the iron-chalcogenide superconductors \cite{bao09}, where the (un)occupied sites in the CDW correspond to up (down) spins in the SDW in those materials. 
When coupling between electronic and lattice degrees of freedom is taken into account, such a picture can also qualitatively explain the distinct peaks in the distribution of bond lengths observed at $T > T_\mathrm{coh}$ via neutron pair distribution function analysis \cite{neilson12}, since links containing a dimer can be expected to be shorter than other links. It can be seen from Figure \ref{cdw_diagram}(b) that each unit cell contains 2 short bonds, 2 long bonds, and 4 bonds of medium length.
%there are an equal number of links containing two spinons (call these ``type A'' links) as there are links containing no spinons (type B), and that the number of links containing just one spinon (type AB) is twice the number of type A or B links. 
This is consistent with the three peaks in the distribution of bond lengths observed in experiment, with the central peak larger than the others. While there is no clear experimental evidence of long-range spatial order such as that described here, we expect that the key features of this model---spatially modulated electron density and dimer formation at high temperatures, giving way to a spatially uniform coherent state at low temperatures---will remain valid even in the absence of long-range order, as illustrated schematically in Figure \ref{cdw_diagram}(c).
%Furthermore, the qualitative results that follow are not dependent on the particular type of charge order that is realized. 

We proceed to study the Hamiltonian \eqref{h} within mean-field theory.
%, assuming that the Coulomb repulsion terms $W_i$ are such that the plaquette order shown in Figure \ref{cdw_diagram}(a) is favored in the CDW phase. 
Denoting as sublattice A (B) the sites shown as (un)occupied in the figure, the average occupation number is taken to be $\left< n_{fi} \right> = n_f + \zeta_i \frac{\Delta}{2}$, where $n_f$ is the average density of spinons per site, $\Delta$ is the CDW order parameter, and $\zeta_i = \pm 1$ on sublattice A (B). 
%In the charge-ordered phase, the unit cell contains 8 sites. 
%Letting the operators $f_1, \ldots, f_8$ correspond to the 8 types of spinon within a unit cell, the Coulomb term in \eqref{h} becomes
%\be
%\label{h_w}
%\begin{split}
%H_W^\mathrm{MF} =& W_3 \sum_{i, \delta''} \left< n_{i + \delta''} \right> 
%	\left( n_i - \frac{1}{2} \left< n_i \right> \right)
%\\	=& W_3 \sum_K \bigg[ 4 \left( n_f - \frac{\Delta}{2} \right) 
%	\left( f^\dagger_{1K} f_{1K} + f^\dagger_{2K} f_{2K} 
%	+ f^\dagger_{3K} f_{3K} + f^\dagger_{4K} f_{4K} \right)
%\\	& \quad \quad \quad \quad + 4 \left( n_f + \frac{\Delta}{2} \right) 
%	\left( f^\dagger_{5K} f_{5K} + f^\dagger_{6K} f_{6K} 
%	+ f^\dagger_{7K} f_{7K} + f^\dagger_{8K} f_{8K} \right) \bigg]
%\\	&\quad \quad + N W_3 ( 4 \Delta^2 - 16 n_f^2 ),
%\end{split}
%\ee
%where $\delta''$ denotes third-nearest neighbors, $K$ is the momentum in the reduced Brillouin zone (the reciprocal space corresponding to unit cells rather than sites) and $N$ is the total number of unit cells. The indices $1, \ldots, 4$ denote the sites that are occupied in Figure \ref{lattice}, while $5, \ldots, 8$ denote those that are unoccupied. 
At mean-field level, the nearest neighbor and second neighbor Coulomb terms $W_{1,2}$ merely shift the chemical potential for the spinons, so they will not be considered further here.
%A convenient choice is
%\be
%\label{zeta}
%\zeta_i = 2 \cos \frac{(2 i_x - 1) \pi}{4} \cos \frac{(2 i_y - 1) \pi}{4}
%\ee
%With the mean-field ansatz \eqref{density}, the Coulomb term in \eqref{h} becomes
%\be
%\label{h_w}
%H_W^\mathrm{MF} = W_3 \sum_k \left[ \Delta (i f^\dagger_{k+Q_1,\sigma} f_{k,\sigma} - f^\dagger_{k+Q_2,\sigma} f_{k,\sigma} + H.c. ) + 4 n_f f^\dagger_{k,\sigma} f_{k,\sigma}+ \frac{\Delta^2}{2} - 2 n_f^2 \right],
%\ee
%where $Q_{1,2} = (\frac{\pi}{2},\pm \frac{\pi}{2})$. 
The Lagrange multiplier field $\lambda_i$ and the slave boson field $b_i$ are also treated as staggered mean fields: $i \lambda_i = \lambda_0 + \zeta_i \lambda_1$ and $b_i = b_0 - \zeta_i b_1$, with $\lambda_{0,1}$ and $b_{0,1}$ real. 
The Heisenberg spins ${\bf S}_i$ are expressed in terms of spinons: $S^a_i = \frac{1}{2} f^\dagger_{i \alpha} \sigma^a_{\alpha \beta} f_{i \beta}$, where $\sigma^a$ are the Pauli matrices. 
%With this, the Heisenberg term in \eqref{h} takes the form
%\be
%\label{hj}
%H_J = - \frac{J}{2} \sum_{\left< ij \right>} 
%	f^\dagger_{i \sigma} f_{j \sigma} f^\dagger_{j \sigma'} f_{i \sigma'}
%	+ J \sum_i n_i - \frac{J}{4} \sum_{\left< ij \right>} n_i n_j.
%\ee
%Since the second and third terms in this equation merely have the effect of shifting the chemical potential and nearest-neighbor Coulomb repulsion, they shall be neglected here. 
We introduce the mean fields $\chi_{A, A', B, AB} = \left< f^\dagger_{i \sigma} f_{j \sigma} \right>$, with $\chi_{AB}$ defined on links between sites on different sublattices, $\chi_B$ on links between two sites on sublattice B, and $\chi_{A (A')}$ on links between two sites on sublattice A in the $x (y)$-direction. 
%While these mean fields will all be equal to one another in the uniform phase, in the CDW phase their values will in general be distinct. 
It is found that the free energy is always lowered in the CDW phase by having only one of $\chi_A, \chi_{A'}$ nonzero, so that the four spins on each plaquette form two dimers. 
In solving the mean-field equations, we require that the average density of spinons is fixed to $n_f = 0.5$ spinons per site in the limit $V=0$, which is accomplished by appropriately setting the on-site energy of the spinons $\varepsilon_f$. 
%This is different from the usual case in theories of condensed matter systems, where the chemical potential is fixed and the particle number is allowed to vary. Accordingly, it is the Helmholtz free energy $F$ rather than the Landau free energy (or ``grand potential'') $\Omega$ that must be minimized, the two quantities being related by $\Omega = F - \mu {\cal N}$, where ${\cal N}$ is the number of particles. Comparing the free energies for the CDW ($\Delta \neq 0$) and normal ($\Delta = 0$) phases allows one to compute the location of the first-order phase transition between the two phases at a given temperature.
%Once the hybridization $V$ is nonzero, $\varepsilon_f$ remains fixed to this value, and in general $n_f \neq 0.5$ once $V$ is finite. 
The chemical potential is set to keep the total density of particles in the system fixed at $n_c + n_f = 1.3$ per site, which remains fixed even for $V \neq 0$.
\begin{figure}
\includegraphics[width=0.45\textwidth]{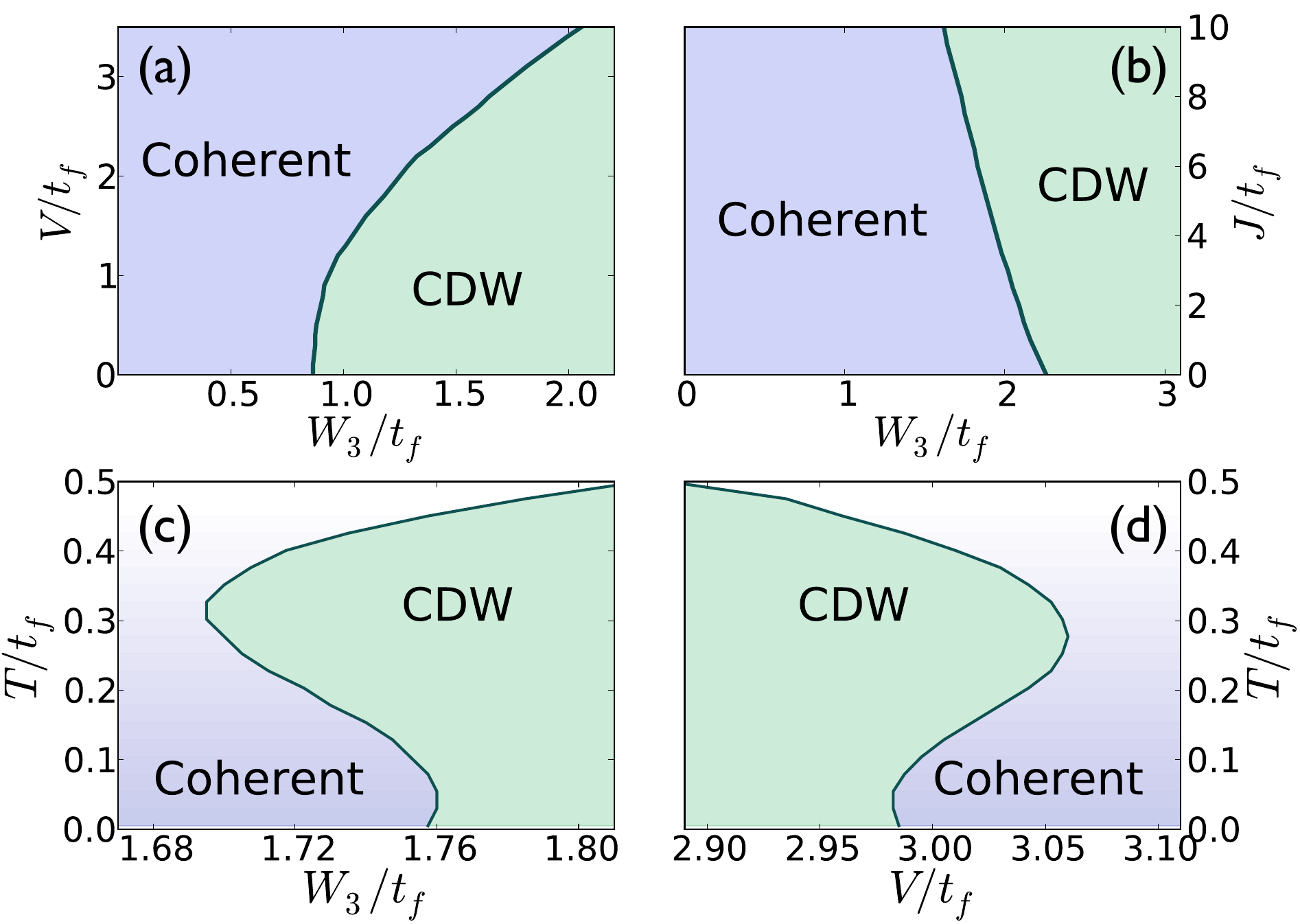}
\caption{Upper plots show the $T=0$ phase diagrams for $V$ vs.\ $W_3$ with $J=8t_f$ (a), and for $J$ vs.\ $W_3$ with $V=3t_f$ (b).
Lower plots show the reentrant transition to the charge-ordered phase as a function of $W_3$ with $V = 3 t_f$ (c), and as a function of $V$ with $W_3 = 1.75 t_f$ (d), and with $J = 8 t_f$ in both plots. All transitions shown are first order.
\label{phase_diagram}}
\end{figure}
The results that follow are not particularly sensitive to the choice of $n_c$, so long as $n_c > n_f$, so that there are enough conduction electrons to screen all of the local moments in the coherent phase. The ratio of hopping amplitudes for spinons and $c$-electrons has been set to $t_f / t_c = 0.2$, which is substantially larger than that found in typical heavy-fermion systems. This is a reflection of the fact that the localized moments in KNi$_2$Se$_2$ are $d$-electrons, which are less tightly bound to their atomic cores than the $f$-electrons that constitute the local moments in most other heavy-fermion systems. This relatively large ratio of bandwidths is also the reason for the rather modest effective mass enhancement of $m^* \sim 10 m_0$ \cite{neilson12}, which is $10 \sim 100$ times smaller than that typically found in $f$-electron heavy fermions.  

The phase diagrams shown in Figure \ref{phase_diagram} illustrate the existence of a reentrant, first-order transition from a CDW to a spatially uniform phase upon cooling. Since the Mermin-Wagner theorem precludes true long-range ordered phases in two dimensions at finite temperature, fluctuations will lead to a phase with CDW correlations but no long-range order, consistent with experimental observations in KNi$_2$Se$_2$. The reentrant behavior is rather unusual, as in most systems the phase that breaks translational symmetry is the ground state that is realized as $T \to 0$. While the reentrance occurs along the entire critical line in Figure \ref{phase_diagram}(a), it emerges along the critical line in Figure \ref{phase_diagram}(b) only for $J \gtrsim 3 t_f$, growing in extent as $J$ is increased. These values of $J$ are rather large to be generated by superexchange alone, for which one expects $J \sim t_f^2/U$. It has been suggested that additional contributions might arise in similar contexts from other superexchange processes in the CDW phase \cite{mckenzie01}, or from RKKY interactions at low temperatures \cite{tran12, coleman89}.
%This shows that the heavy quasiparticles are able to reduce their energy by entering a spatially uniform, correlated phase as $T$ is lowered.  
Reentrant behavior reminiscent of that shown here has been seen previously in a theory of simple checkerboard CDW ordering at $1/4$ filling in layered molecular crystals \cite{merino01}, although in that case only a second-order transition was found. The model presented here also exhibits a second-order transition, but only at higher temperatures than those shown in Figure \ref{phase_diagram}. The mean field $\chi_A$ is nonzero throughout the CDW phase shown in Figure \ref{phase_diagram}, indicating dimer formation between spinons. 

%As we mentioned above, this singlet formation has important experimental consequences. One such consequence, to be discussed in more detail below, is the lack of Curie magnetic response from the local moments, which is typically observed at high temperatures in heavy-fermion systems. Another is the absence of a resistivity peak as temperature is lowered. Such a peak typically occurs in heavy-fermion systems below the single-ion Kondo temperature, where conduction electrons begin to screen the local moments, but above the temperature at which these resonances become coherent, forming a heavy-fermion state. It is the strong spin-flip scattering from these resonances that typically gives rise to the resistivity peak in Kondo systems. The absence of such a peak \cite{neilson12} in KNi$_2$Se$_2$ indicates that there is no intermediate single-ion Kondo regime in this material. Instead of forming Kondo screening clouds that act as strong scattering centers, the local moments form dimers in the CDW phase, and then this gives way directly to a coherent heavy-fermion phase, with no intermediate single-ion Kondo regime. 

For the parameters given above, there is a jump in the average spinon occupation per site $n_f$ at the first-order transition. At $W_3 = 1.75 t_f$, the occupation jumps from $n_f = 0.45$ for $T < T_\mathrm{coh} = 0.12 t_f$ to $n_f = 0.53$ for $T > T_\mathrm{coh} $. 
%The reduced valency $n_f < 0.5$ in the uniform phase is expected, indicating that some of the spinons become delocalized and join the sea of conduction electrons in forming a heavy Fermi liquid. 
The increased valency in the CDW phase is consistent with the lack of long-range CDW order observed in experiment \cite{neilson12}, since a long-range ordered state would be impossible at incommensurate filling.
%, which is given by $n_f$ on account of the constraint $\langle n_{fi} \rangle + b_i^2 = 1$, 
It would be interesting to test whether the predicted jump in $n_f$ at $T = T_\mathrm{coh}$ could be observed experimentally using techniques such as resonant inelastic X-ray scattering \cite{dallera02}.

\begin{figure}
\includegraphics[width=0.45\textwidth]{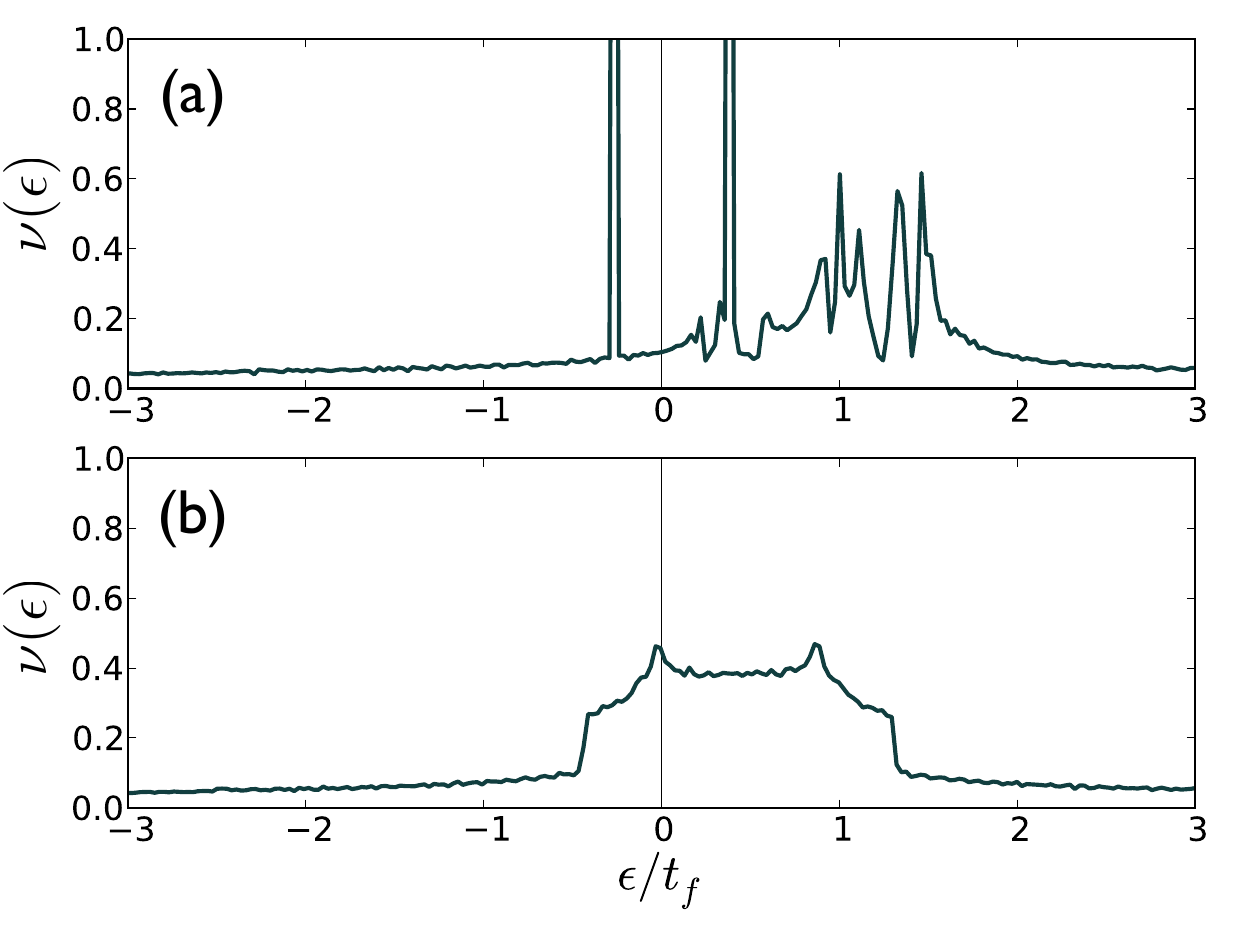}
\caption{Densities of states for the CDW phase at $T = 0.15 t_f$ (a) and the coherent phase at $T = 0.01 t_f$ (b), with $W_3 = 1.75 t_f$, $J = 8 t_f$ and $V = 3 t_f$. The Fermi level is at $\epsilon = 0$.
\label{dos}}
\end{figure}

The densities of states in the two phases are shown in Figure \ref{dos}. 
In the CDW phase, the well-defined peaks above and below the Fermi level clearly show that the spinon excitations are gapped. In contrast, the spinons contribute to the hybridization peak at the Fermi level in the low-temperature phase, as is typical in heavy-fermion materials. The relative magnitude of the peak in this case, however, is substantially smaller than that in $f$-electron materials. Comparing the values of the densities of states at the Fermi level in the two different phases, one finds with these particular parameters an enhancement of $\approx 3.0$ in the normal phase relative to the CDW phase. This can be compared with measurements on KNi$_2$Se$_2$, where an enhancement of \hbox{$\approx 3.1$} was observed in the electronic specific heat coefficient $\gamma$ for $T < T_\mathrm{coh}$ \cite{neilson12}.

The uniform magnetic susceptibility can be calculated as the derivative of magnetization with respect to applied magnetic field.
%\be
%\chi (T) = \lim_{B \to 0} \frac{dm}{dB},
%\ee
%where $m$ is the magnetization and $B$ is the applied magnetic field.
\begin{figure}
\includegraphics[width=0.45\textwidth]{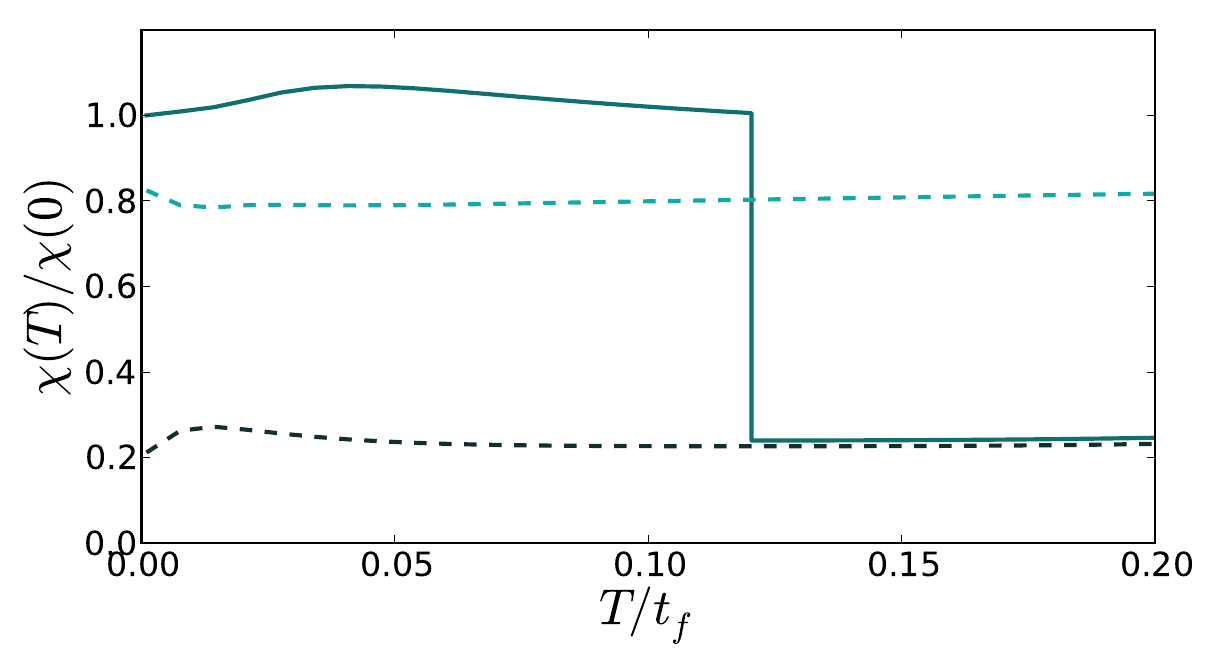}
\caption{Uniform magnetic susceptibility with $W_3 = 1.75 t_f$ and $J = 8 t_f$. The light and dark dashed lines show $\chi(T)$ in the uniform phase ($V = 4 t_f$) and in the CDW phase ($V = 2 t_f$), respectively. The solid line shows $\chi (T)$ for $V = 3 t_f$, which exhibits a phase transition at $T = T_\mathrm{coh} = 0.12 t_f$. All plots are normalized to $\chi (0)$ for the solid curve.
\label{susceptibility}}
\end{figure}
Figure \ref{susceptibility} shows the susceptibility as a function of temperature, assuming that the Lande $g$-factors for conduction electrons and spinons are equal. 
%A low-temperature peak is followed by a region of approximately constant susceptibility at intermediate temperatures. This agrees with the experimental observation \cite{neilson12}, where a similar low-temperature peak was observed. 
The approximately constant susceptibility in the region $T < 0.12 t_f = T_\mathrm{coh}$ corresponds to the Pauli susceptibility of the heavy Fermi liquid, in which the conduction electrons are hybridized with the localized spinons, leading to an enhanced density of states near the Fermi level. At $T_\mathrm{coh}$, there is a jump in the susceptibility to a much smaller constant value, indicating that only the conduction electrons contribute to the susceptibility at $T > T_\mathrm{coh}$, while the spinons form singlet pairs. This is in contrast to the Curie susceptibility $\chi \sim 1/T$ that is typically observed at high temperatures in heavy-fermion materials. 
%The ratio of susceptibilities at temperatures just above and below $T = T_\mathrm{coh}$ is $\sim 4$. 

Rather than exhibiting a sharp step, however, the experimentally measured $\chi(T)$ remains approximately constant through the CDW transition \cite{neilson12}. One possible explanation for this discrepancy is the Van Vleck contribution to $\chi (T)$, which has not been included in our model. The possibility of a large contribution of this type in heavy-fermion materials has been considered previously \cite{zou86}. Others have since investigated the Van Vleck contribution to the susceptibility and have found that, when multiple localized bands are approximately degenerate, one generally has $\chi_V \sim \chi_\mathrm{Pauli}$ \cite{evans90, kontani96, mutou96}. 
%The Van Vleck contribution arises from the Zeeman coupling at second-order in perturbation theory:
%\be
%\label{chi_v}
%\chi_V = 2 \mu^2 \sum_{\gamma \neq \gamma_0} 
%	\bigg< \frac{| \langle \gamma | L^z + 2 S^z | \gamma_0 \rangle | ^2}{E_{\gamma_0} - E_\gamma} \bigg>_\mathrm{FS},
%\ee
%where $\gamma$ denotes electron orbitals, $\langle \cdots \rangle_\mathrm{FS}$ denotes an average over the Fermi surface, and the magnetic field is assumed to be oriented along the $z$-direction. The inter-orbital energy splitting in the denominator of \eqref{chi_v} allows for $\chi_V$ to become large if this splitting is small. If two or more of the $d$-electron bands are nearly degenerate, then only one of these bands will hybridize with the conduction electrons. If the splitting is smaller in the CDW phase than in the low-temperature coherent phase (\textit{i.e.}\ if the unhybridized band has large spectral weight near one of the peaks in Figure \ref{dos}(a)), then this may provide a mechanism that leads to a susceptibility that remains roughly constant across the CDW transition. Since the Pauli susceptibility is much larger in typical heavy-fermion systems than in KNi$_2$Se$_2$, the Van Vleck term can be expected to provide a greater relative contribution in the latter case. 
Thus if an unhybridized band that was not included in our model has large spectral weight near one of the peaks in Figure \ref{dos}(a), an increased $\chi_V$ could compensate for the decrease in $\chi_\mathrm{Pauli}$ at higher temperatures, with the sum of the two terms remaining roughly constant.  Calculating the precise Van Vleck contribution to the susceptibility would require a more detailed knowledge of the band structure, however, and so we leave this as an open question to be addressed in future work. 

In conclusion, we have provided a theoretical framework for describing the key properties of the recently discovered mixed-valency material KNi$_2$Se$_2$, most importantly the vanishing of the CDW phase upon cooling. The formation of singlet dimers by the local moments in the CDW phase explains the lack of common signatures of single-ion Kondo behavior, such as a Curie susceptibility at high temperatures. This mechanism may also explain the lack of a resistivity peak in measurements on KNi$_2$Se$_2$ \cite{neilson12}. Such a peak typically forms in heavy-fermion materials at temperatures just above the coherence temperature, where the Kondo screening clouds are not yet coherent with one another and act as spin-flip scattering centers. A direct transition from a singlet CDW phase to a coherent low-temperature phase precludes this possibility, however, and is consistent with the monotonically decreasing resistivity observed in experiment as $T$ is lowered. This material illustrates the potential of mixed-valency systems for exhibiting a rich array of collective quantum behaviors.

We thank J.\ Kang, T.\ McQueen, J.\ Neilson, O.\ Tchernyshyov, and Y.\ Wan for helpful discussions. This work was supported by the Johns Hopkins--Princeton Institute for Quantum Matter, under Grant No.\ DE-FG02-08ER46544 from the US Department of Energy, Office of Basic Energy Sciences, Division of Materials Sciences and Engineering.

\bibliographystyle{apsrev}

\end {document}